\documentclass[a4paper,10pt]{aastex62}

\hypersetup{linkcolor=red,citecolor=blue,filecolor=cyan,urlcolor=magenta}
\usepackage{natbib}
\usepackage{amsthm}
\usepackage{amssymb}
\usepackage{amsmath}
\usepackage{bm}
\usepackage{graphicx}
\usepackage[caption=false]{subfig}

\shorttitle{PIC simulations of the Firehose Instabilities}
\shortauthors{Micera et al.}

\begin{document}

\title{Particle-In-Cell simulations of the parallel proton firehose instability influenced by the electron temperature anisotropy in solar wind conditions}

\correspondingauthor{A. Micera}
\email{alfredo.micera@oma.be}

\author{A. Micera}
\affil{Solar-Terrestrial Centre of Excellence - SIDC, Royal Observatory of Belgium, Brussels, Belgium.}
\affil{Centre for Mathematical Plasma Astrophysics, KU Leuven, Leuven, Belgium.}

\author{E. Boella}
\affiliation{Physics Department, Lancaster University, Lancaster, UK.}
\affiliation{Cockcroft Institute, Daresbury Laboratory, Warrington, UK.}

\author{A. N. Zhukov}
\affiliation{Solar-Terrestrial Centre of Excellence - SIDC, Royal Observatory of Belgium, Brussels, Belgium.}
\affiliation{Skobeltsyn Institute of Nuclear Physics, Moscow State University, Moscow, Russia.}

\author{S. M. Shaaban}
\affiliation{Centre for Mathematical Plasma Astrophysics, KU Leuven, Leuven, Belgium.}
\affiliation{Theoretical Physics Research Group, Physics Department, Mansoura University, Mansoura, Egypt.}

\author{R. A. L\'{o}pez}
\affiliation{Centre for Mathematical Plasma Astrophysics, KU Leuven, Leuven, Belgium.}

\author{M. Lazar}
\affiliation{Centre for Mathematical Plasma Astrophysics, KU Leuven, Leuven, Belgium.}
\affiliation{Institut f\"{u}r Theoretische Physik, Weltraum- und Astrophysik, Ruhr-Universit\"{a}t Bochum, Bochum, Germany.}

\author{G. Lapenta}
\affiliation{Centre for Mathematical Plasma Astrophysics, KU Leuven, Leuven, Belgium.}

\begin{abstract}
\textit{In situ} observations of the solar wind show a limited level of particle temperature anisotropy with respect to the interplanetary magnetic field direction. Kinetic electromagnetic instabilities are efficient to prevent the excessive growth of the anisotropy of particle velocity distribution functions. Among them, the firehose instabilities are often considered to prevent the increase of the parallel temperature and hence to shape the velocity distribution functions of electrons and protons in the solar wind. We present a non-linear modeling of the parallel firehose instability, retaining a kinetic description for both the electrons and protons. One-dimensional (1D) fully kinetic Particle-In-Cell simulations using the Energy Conserving semi-implicit method (ECsim) are performed to clarify the role of the electron temperature anisotropy in the development of the parallel proton firehose instability.  
We found that in the presence of an electron temperature anisotropy, such that the temperature parallel to the background magnetic field is higher than the temperature in the perpendicular direction, the onset of the parallel proton firehose instability occurs earlier and its growth rate is faster. 
The enhanced wave fluctuations contribute to the particle scattering reducing the temperature anisotropy to a stable, nearly isotropic state. The simulation results compare well with linear theory. A test case of 1D simulations at oblique angles with respect to the magnetic field is also considered, as a first step to study the cumulative effect of protons and electrons on the full spectrum of instabilities.

\end{abstract}

\keywords{plasmas --- solar wind --- instabilities --- waves }
%
\section{Introduction}\label{sec.0}
%
Solar wind is a natural plasma laboratory, enabling \textit{in situ} measurements and investigations of a rich variety of physical processes operating there. More than 50 years of observations (made e.g. by \textit{Helios}, \textit{Ulysses}, and \textit{Wind} space missions) revealed that proton and electron velocity distribution functions (VDFs) may be markedly non-Maxwellian, and therefore non-thermal. The observed distributions are often anisotropic with respect to the magnetic field direction \citep[see][and references therein]{Marsch1982, Marsch2006, Stverak2008, Kasper2016}. Deviations from isotropy are limited by such mechanisms as particle-particle (binary) collisions and wave-particle interactions, resulting in fluctuations of electric and magnetic fields which scatter plasma particles and reduce their temperature anisotropy. The fluctuations may be either pre-existing (e.g. turbulence) or generated by plasma instabilities. These mechanisms are usually invoked in the models to explain the observed VDFs \citep{Hellinger2006, Stverak2008, Bale2009, Chen2016, Shaaban2017, Lazar2017, Shaaban2019a, Shaaban2019b, Ofman2019}.

Solar wind is a hot and low-density plasma, such that binary collisions are not efficient to limit the temperature anisotropies of plasma particles. Despite significant efforts in fluid modeling \citep[see e.g. the review by][]{Hunana2019}, a kinetic approach is needed to describe other mechanisms potentially constraining the anisotropies, e.g., the plasma kinetic instabilities and wave-particle interactions, and to explain the low level of anisotropy recorded \textit{in situ} for both electrons and protons at 1 AU \citep{Matteini2007, Stverak2008}. A kinetic approach is also important to take into account wave-like processes operating at frequencies ranging between the ion gyrofrequency and the electron plasma frequency \citep{Bruno2013}. Spectra of wave-like fluctuations of the solar wind electric and magnetic fields are well established by the observations, including those at kinetic scales. These waves can resonantly interact with particles resulting in the exchange of energy and momentum between them \citep{Alexandrova2013}. These are key aspects to understand the kinetic processes in the solar corona and the solar wind. 

Kinetic simulations enable detailed studies of the plasma instabilities and their consequences, e.g., enhancement of the observed wave fluctuations and their reaction on particles that may lead to their energization. They allow for modelling the plasma from first principles and can correctly describe non-thermal behaviours, anisotropic distribution functions and non-symmetric pressure tensors in the solar wind plasma. Among the different numerical techniques, the Particle-In-Cell (PIC) method \citep{Birdsall_Langdon, Hockney} represents our choice because the algorithm is scalable and can take advantage of modern high-performance computing systems \citep{Gonzalez-Herrero2018}.

Firehose instabilities are thought to be a crucial mechanism for constraining the electron and proton temperature anisotropies observed in expanding astrophysical plasmas, such as the solar wind \citep{Bale2009, Chen2016, Lazar2017, Shaaban2019a}. These electromagnetic kinetic instabilities are thought to develop around 1 AU and beyond \citep{Matteini2006}, but perhaps also closer to the Sun \citep{VanderHolst2019}.
Firehose instabilities can develop in a plasma with sufficiently large parallel plasma beta $\beta_{j, \parallel} = 8 \pi n_j k_B T_{j, \parallel} / B_0^2 >~ 1$ and anisotropy $A_j = T_{j, \perp} / T_{j, \parallel} < 1$, where $j$ refers to a given plasma species ($j=e,p$ for electrons and protons), $n$ and $T$ are the particle number density and temperature respectively, $k_{B}$ the Boltzmann constant, while $\perp$ and $\parallel$ denote directions perpendicular and parallel to the background magnetic field, $\bm{B}_0$.

A number of studies have been conducted to demonstrate the co-existence of the firehose instabilities with evolving turbulent spectra \citep{Hellingeretal2015}, to confirm the evolution of Alfv{\'e}nic fluctuations in the firehose regime \citep{TeneraniVelli2018} and to test the properties of the instabilities in the expanding solar wind \citep{Hellingeretal2003, Matteini2006, Innocenti2019,Innocentietal2019,Hellingeretal2019}. 
The proton firehose instability has been largely examined via hybrid simulations, e.g. \citep{Matteini2006, Hellinger2017}, where the interest is focused only on the ion-scale processes, since the electrons are considered as a massless, charge neutralizing fluid. The proton firehose instability evolves out of the whistler wave and has right-hand circular polarization \citep{Gary1993_book}.

Two branches of the firehose instability can be driven by the anisotropic proton distributions: the periodic proton firehose with the finite real part of the wave frequency $\omega_r \ne 0$ and the most unstable modes in the direction parallel to the magnetic field, and the aperiodic modes ($\omega_r = 0$) that are present only for oblique angles of propagation, \cite[e.g.][]{Yoon1993, HellingerMatsumoto2000, Hunana2017}. 
The properties of these two modes, when predicted using simplified models neglecting the effects of electrons, do not always agree with observations and contradictory conclusions were sometimes obtained. For example, \citet{Kasper2002} compared observations of proton velocity distribution functions with the limits for parallel and oblique firehose instabilities derived from the linear theory and simulations as reported by \citet{Gary1998}. The agreement with the parallel mode was better.
\citet{HellingerMatsumoto2000} found that the oblique proton firehose instability may have the growth rate comparable or even grater than that of the parallel proton firehose.
\citet{HellingerMatsumoto2001} extended the work by \citet{HellingerMatsumoto2000} and determined that the parallel firehose instability initially dominates and saturates faster than the oblique one, which then can grow even when the parallel modes saturate.
\citet{Hellingeretal2016} and \citet{Bale2009} state that the periodic (parallel) mode should develop faster, but the observed limits of temperature anisotropy are better shaped by the thresholds of the aperiodic (oblique) firehose.
However, \citet{Hellinger2006} analyzed proton firehose instabilities in the presence of alpha particles and concluded that the parallel and oblique modes have comparable growth rates for a vast range of plasma parameters \citep[see also][]{Hunana2017,Hunana2019} and that both modes are relevant in the solar wind context.

On the other hand, the observations confirm the (co-)existence of solar wind electrons with $T_{e\parallel} > T_{e\perp}$ \citep{Marsch2006,Stverak2008}, which can trigger two similar branches of the electron firehose instability. There is no unique definition for the two branches of the electron firehose instability. \citet{LiandHabbal2000} define the quasi-parallel branch as periodic modes in directions within 30$^\circ$ with respect to the magnetic field, while the oblique branch represents aperiodic modes at larger angles. \citet{GaryandNashimura2003} define the parallel electron firehose as non-resonant while the oblique mode is resonant. \citet{Camporeale2008} distinguish propagating ($\omega_r \neq 0$) and non-propagating ($\omega_r = 0$) modes.  
Even though \citet{Camporeale2008}, in agreement with \citet{LiandHabbal2000}, did not find any non-propagating perturbation at small angles, they demonstrated the extension of propagating perturbations to relatively high oblique angles.
Both branches have been investigated but generally this has been done neglecting their interplay with the proton firehose instability \citep{QuestShapiro1996, Paesold1999, LiandHabbal2000, GaryandNashimura2003,Paesold2003,Camporeale2008, Hellingeretal2014, Shaaban2019c, Lopez2019}. Neglecting the interplay of different species may prevent understanding of the instability implications and the anisotropy relaxation mechanisms \citep{Yoon2017b,Yoon2019}.

Recently, more realistic approaches have been proposed showing that linear properties of both branches can be markedly altered by the interplay of anisotropic electrons and protons \citep{Michno2014, Maneva2016, Shaaban2017} and may thus lead to a better agreement with observations. However, these approaches do not allow analyzing such phenomena as the effect of the wave growth on the scattering of particles and hence on the rate of their isotropization.
In an attempt to check and complete these predictions from linear theory, in this work we perform one-dimensional (1D) PIC simulations to investigate the periodic parallel branch of the proton firehose instability under the effect of anisotropic electrons. We are aware that reducing the scope of our investigation to perturbations propagating only in the parallel direction may be a limitation. The oblique electron firehose instability has generally a lower threshold and higher growth rate than the parallel electron firehose instability. Nevertheless, quasi-parallel and exactly parallel modes may be dominant after the saturation of the linear stage \citep{Camporeale2008,Innocenti2019}. Even if our work does not deal with the competition between parallel and oblique modes, it is the first numerical study that looks at the nonlinear evolution of the parallel proton firehose instability in the presence of anisotropic electrons using fully kinetic simulations, which retain a description from first principles for both electrons and ions.

This paper is organized as follows. Section \ref{sec.2} illustrates the employed simulation setup. Section \ref{seccc.2} reports the main results of our simulations. In Section \ref{sec.3} simulation results are compared with the linear theory. Section \ref{sec.4} presents the results of a few tests of simulations at oblique angles with respect to the magnetic field. Finally, Section \ref{sec.5} presents a discussion of the simulation results in the solar wind context, and reports the conclusions.

%
\section{Setup of the PIC simulations}\label{sec.2}
%
In order to study the development and evolution of the parallel firehose instability in the solar wind, we performed 1D kinetic simulations using the ECsim code \citep[Energy Conserving semi-implicit method, see][]{Lapenta2017,Lapentaetal2017,Gonzalez-Herrero2018, Gonzalez-Herrero2019}. ECsim is a semi-implicit PIC code which, unlike hybrid methods, is capable to retain kinetic electron and ion information, such as wave-particle interaction and non-Maxwellian VDFs. It has been demonstrated that the code is stable and energy conserving over a wide range of spatial and temporal resolutions \citep{Lapentaetal2017, Gonzalez-Herrero2018}.
In particular, the latter aspect is crucial for the current work, because it allows us to resolve only the scales of interest. This means that we do not need to resolve the electron Debye length as in traditional PIC codes \citep{Cohen1981, Brackbill1982,Lapenta2017}. As a consequence, we can follow the dynamics of both electrons and protons from a kinetic point of view for very long times, thus exploring the interplay of electrons and protons, and the long-term evolution of the firehose instability.

We model a collisionless plasma composed of electrons and protons with real mass ratio $\mu =~m_p / m_e =~1836$, immersed in a uniform background magnetic field $\bm{B}_0$. The magnetic field is along the $x$-direction ($\bm{B}_0=B_0 \hat{e}_x$), and its magnitude $B_0 = 2.5\ 10^{-4}$~G  is such that $\omega_e / \Omega_e = 63.24$ and $\omega_p / \Omega_p = 2709.98$, with $\omega_j = \sqrt{4 \pi e^2 n_j / m_j}$ and $\Omega_j = \sqrt{e B_0 / c\, m_j}$ being the plasma and the cyclotron frequencies (for species $j$), respectively, $e$ is the elementary charge, $m$ is the species mass and $c$ is the speed of the light in vacuum. The initial values of the electron number density ($n = 25$~cm$^{-3}$), magnetic field magnitude and $\omega_{e} / \Omega_{e}$ are typical for the solar wind around 0.3 -- 0.5 AU, \cite[e.g.][]{Bothmer2019,Tong2019}.
A simulation box with the length of $60\, d_i$ has been employed, where $d_i = c / \omega_p$ is the ion inertial length. The size of the box is chosen so that more than $20$ wavelengths of the most unstable mode fit into the box. A cell size $\Delta x \simeq 0.074\, d_i$ and a temporal step $\Delta t = 0.5\, \omega_p^{-1}$ have been chosen. We therefore fully resolve the ion inertial length, and characteristic electron scales are resolved only marginally. This is sufficient as the electron firehose instability occurs at proton scales \citep{Gary1998,LiandHabbal2000}. Both species are initially described by bi-Maxwellian velocity distribution functions with no drift velocity:

\begin{equation}
     \begin{aligned}
F_j (v_{\parallel}, v_{\perp}) = &\frac{1}{\pi^{3/2} u^2_{j, \perp} u_{j, \parallel}} \exp \left( -\frac{v_{\parallel}^2}{u_{j, \parallel}^2 }  -\frac{v_{\perp}^2}{u_{j, \perp}^2} \right), 
    \label{sss}   
     \end{aligned}     
\end{equation}
with $u_j = \sqrt{2 k_B T_j / m_j}$ is the thermal velocity of the species $j$. 

In all the simulations we use $10^4$ particles per cell per species, and adopt periodic boundary conditions for fields and particles. Simulations performed with different resolution and number of particles per cell yield similar results, showing a good convergence of the code. This convergence study together with the possibility of easily comparing the numerical outcomes with the linear theory of the periodic proton firehose instability, led us to infer that the code is correct and accurate, even when electron scales are resolved only marginally.

%
\section{Results of the PIC simulations}\label{seccc.2}
%
In order to determine the role of the interplay of electrons and protons in the firehose instability, we compare four cases: the pure proton firehose instability (PFHI) where only the protons are anisotropic (case 1), the PFHI where electrons are also anisotropic but characterized by low plasma beta (case 2), the pure electron firehose instability (EFHI) where only the electrons are anisotropic (case 3), and
the case characterized by the temperature anisotropy of both species (PFHI + EFHI) and high electron plasma beta (case 4), see Table~\ref{table:nonlin}.
The values of initial anisotropies $A_j=T_{j,\perp}/T_{j,\parallel}$ are perhaps lower with respect to the average solar wind conditions. This choice is mainly motivated by simulation reasons, where the computational resources constitute a serious constraint. The simulations with higher values of $A_j$ are characterized by lower growth rates and hence they run significantly longer before achieving the saturation. All the other parameters such as electron and proton number densities, magnitude of the magnetic field, ratio between the electron plasma frequency and electron cyclotron frequency are typical for the solar wind as mentioned in Section \ref{sec.2}. Finally we note that the values of anisotropy and plasma beta in the solar wind are affected by large fluctuations with values that approach those in our simulations \citep[e.g.][]{kasper2003, Hellingeretal2016, Bale2009}.

\begin{table}[ht]
\caption{Input parameters}
\centering
\begin{tabular}{l c c c c}
\hline\hline
 Case &$\beta_{p, \parallel}$&$\beta_{e, \parallel}$&$A_{p}$&$A_e$ \\ [0.5ex]
\hline
 1 (PFHI) & 4.0 & 4.0 & 0.1 & 1.0  \\
 2 (PFHI) & 4.0 & 0.1 & 0.1 & 0.1  \\
 3 (EFHI) & 4.0 & 4.0 & 1.0 & 0.1  \\
 4 (PFHI + EFHI) & 4.0 & 4.0 & 0.1 & 0.1 \\ [1ex] 
\hline
\end{tabular}
\label{table:nonlin}
\end{table}

\begin{figure}
\centering
\includegraphics[scale=0.22]{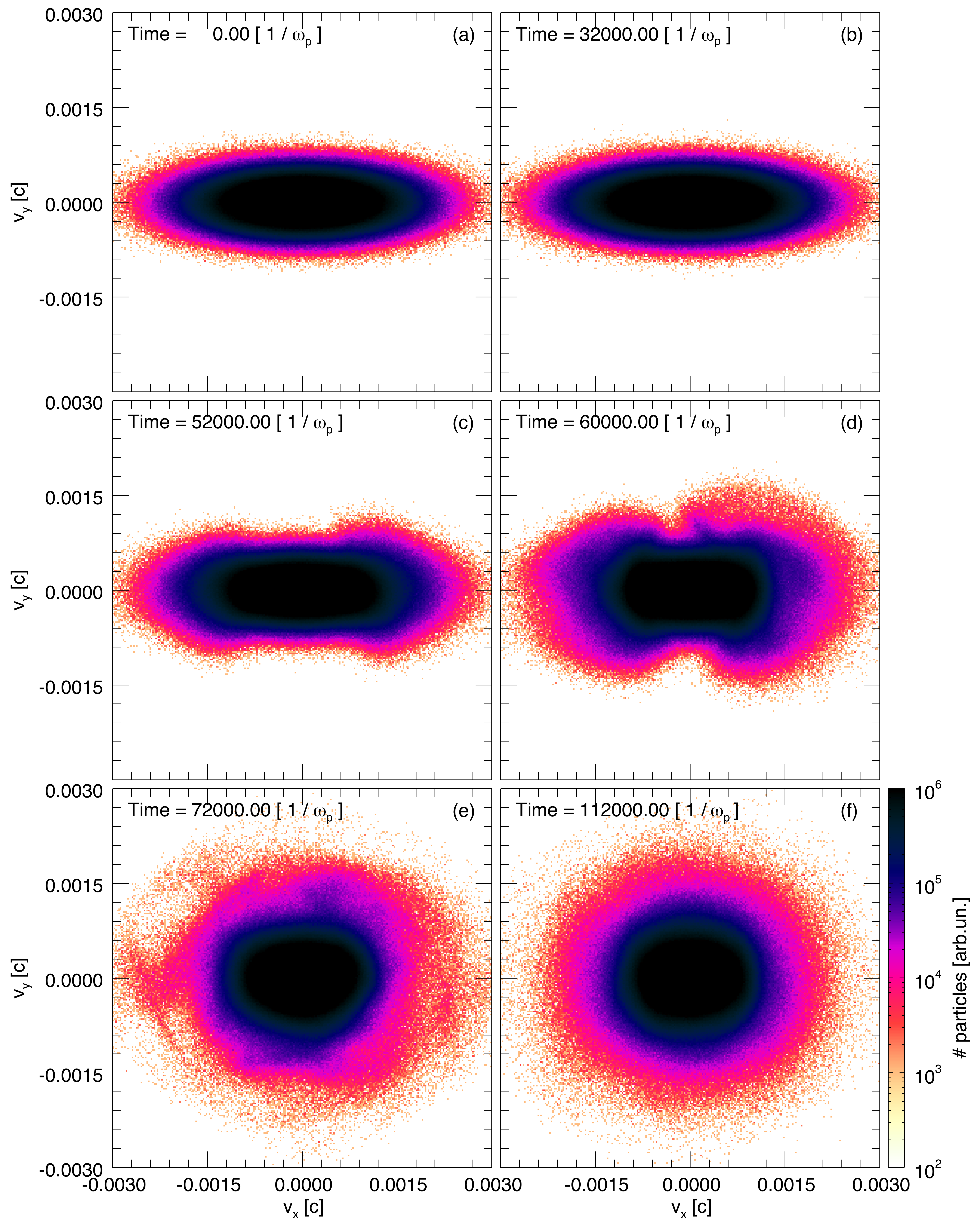}
\caption{Proton VDFs in the PFHI case (case 1 of Table \ref{table:nonlin}) at $t = 0$ (a), $32000$ (b), $52000$ (c), $60000$ (d), $72000$ (e), $112000$ (f). In all the figures of the paper, the time is in units of $\omega_p^{-1}$, the velocities are in units of $c$ and the number of particles is in arbitrary units. The background magnetic field is directed parallel to the $x$ axis.}\label{fig.3}
\end{figure}

In Figure \ref{fig.3} the evolution of the proton velocity distribution function is displayed for the PFHI case 1. Proton VDFs are shown at the initial time step (Figure \ref{fig.3}(a)), during the development of the instability (Figure \ref{fig.3}(b)-(e)) and after the instability saturation (Figure \ref{fig.3}(f)).
One can notice that protons are scattered towards higher perpendicular velocities by resonant wave-particle interactions. The process continues until a condition of marginal stability is reached. The isotropic species, i.e. the electrons, does not play any particular role in the development of the instability. The electron temperature is nearly constant during the simulation and their VDF remains Maxwellian. 

Case 2 of Table \ref{table:nonlin} is characterized by both species anisotropic but with electron plasma beta too low to excite the electron firehose instability. Similarly to case 1, only the proton firehose instability develops, and the evolution of the proton VDF is very similar to that shown in Figure \ref{fig.3} so it is not shown here. 

\begin{figure}[] 
\centering
\includegraphics[scale=0.22]{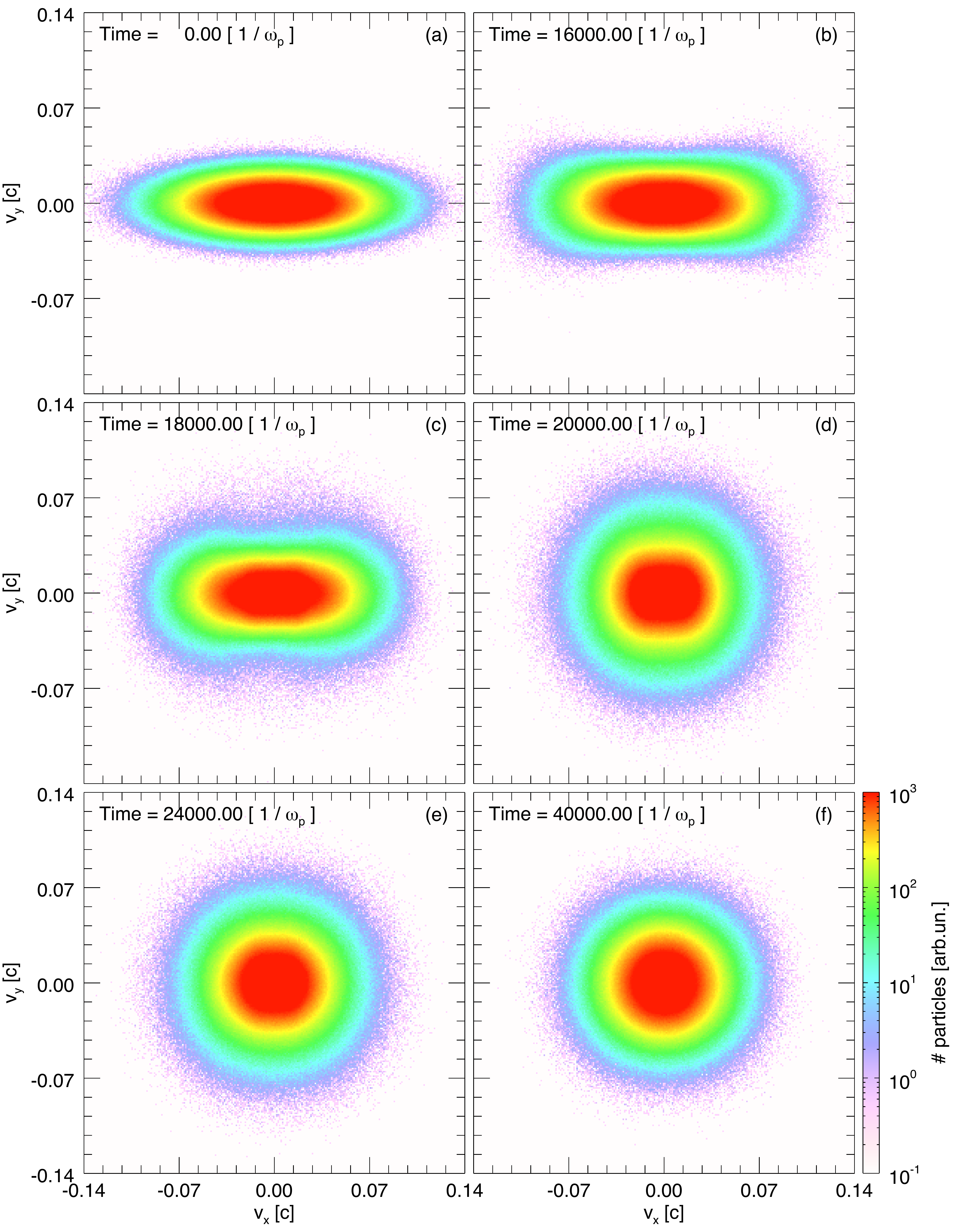}
\caption{Electron VDFs in the EFHI case 3 of Table \ref{table:nonlin} at $t = 0$ (a), $16000$ (b), $18000$ (c), $20000$ (d), $24000$ (e), $40000$ (f).}\label{figg.5}
\end{figure}

Figure \ref{figg.5} shows the evolution of the electron VDF in the EFHI case 3. The anisotropy triggers the instability and the free energy of the particles is converted into magnetic energy. The anisotropy is reduced during the evolution of the instability through the scattering between particles and electromagnetic waves (Figure \ref{figg.5}(b)-(e)) until a complete isotropization (Figure \ref{figg.5}(f)). 
Similarly to case 1, the isotropic species, the protons, does not participate to the development of the instability.
If one compares Figure \ref{fig.3} and Figure \ref{figg.5}, i.e. the evolution of the PFHI with the evolution of the EFHI, it can be clearly seen that the latter develops on shorter time scales than the former, so the process of wave-particle interactions is more efficient and rapid in isotropizing the anisotropic species.
Electrons get heated in the perpendicular direction mainly through non-resonant wave-particle interaction, as it should be for the parallel EFHI, which is usually not resonant. As mentioned by \citet{Messmer2002}, the non-resonant wave-particle interaction results in electrons with no preferential $v_{\parallel}$ being pitch-angle scattered. This results in an increase of $v_{\bot}$. However, the electron VDF at $t = 18000 \, \omega_p^{-1}$ has the characteristic butterfly shape and shows a slight signature of a resonance. This is certainly due to the high level of anisotropy. The EFHI becomes indeed resonant with the electrons for high anisotropy values \citep{Paesold1999}.

\begin{figure}
\centering
\includegraphics[scale=0.20]{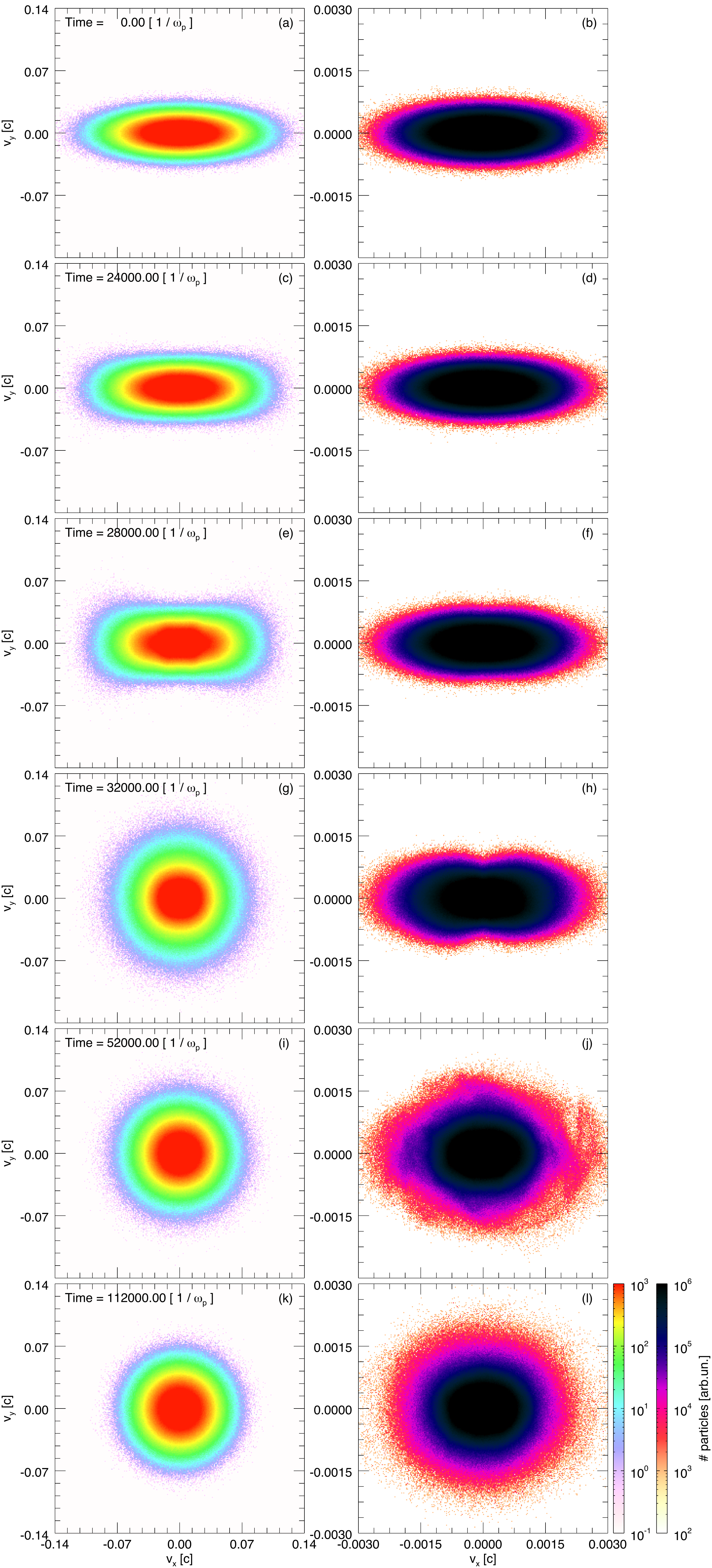}
\caption{Electron (left) and proton (right) VDFs in the PFHI + EFHI  case 4 of Table \ref{table:nonlin} at $t = 0$ (a, b), $24000$ (c, d), $28000$ (e, f), $32000$ (g, h), $52000$ (i, j), $112000$ (k, l).}\label{fig.4}
\end{figure}

Figure \ref{fig.4} reports the evolution of the electron and proton VDFs in the PFHI + EFHI  case 4. The particle VDFs at $t = 0$ (Figure \ref{fig.4}(a)-(b)), during the instability (Figure \ref{fig.4}(c)-(f), (h) and (j)), and at the instability saturation (Figure \ref{fig.4}(g), (i), (k) and (l)) are shown.
Electrons start to isotropize much earlier than protons (Figure~\ref{fig.4}(c)) and absorb a part of the free energy before wave-proton interaction takes place. Once electrons reach the condition of marginal 
stability, protons start to be scattered by the interaction with electromagnetic waves until their isotropization.
Comparing Figure \ref{fig.3} with Figure \ref{fig.4}, one can note that in  the PFHI + EFHI  simulation, where both species are initially anisotropic, the exchange of energy between protons and the waves starts earlier than in the PFHI case where only protons are anisotropic. 

\begin{figure}
\centering
\includegraphics[scale=0.22]{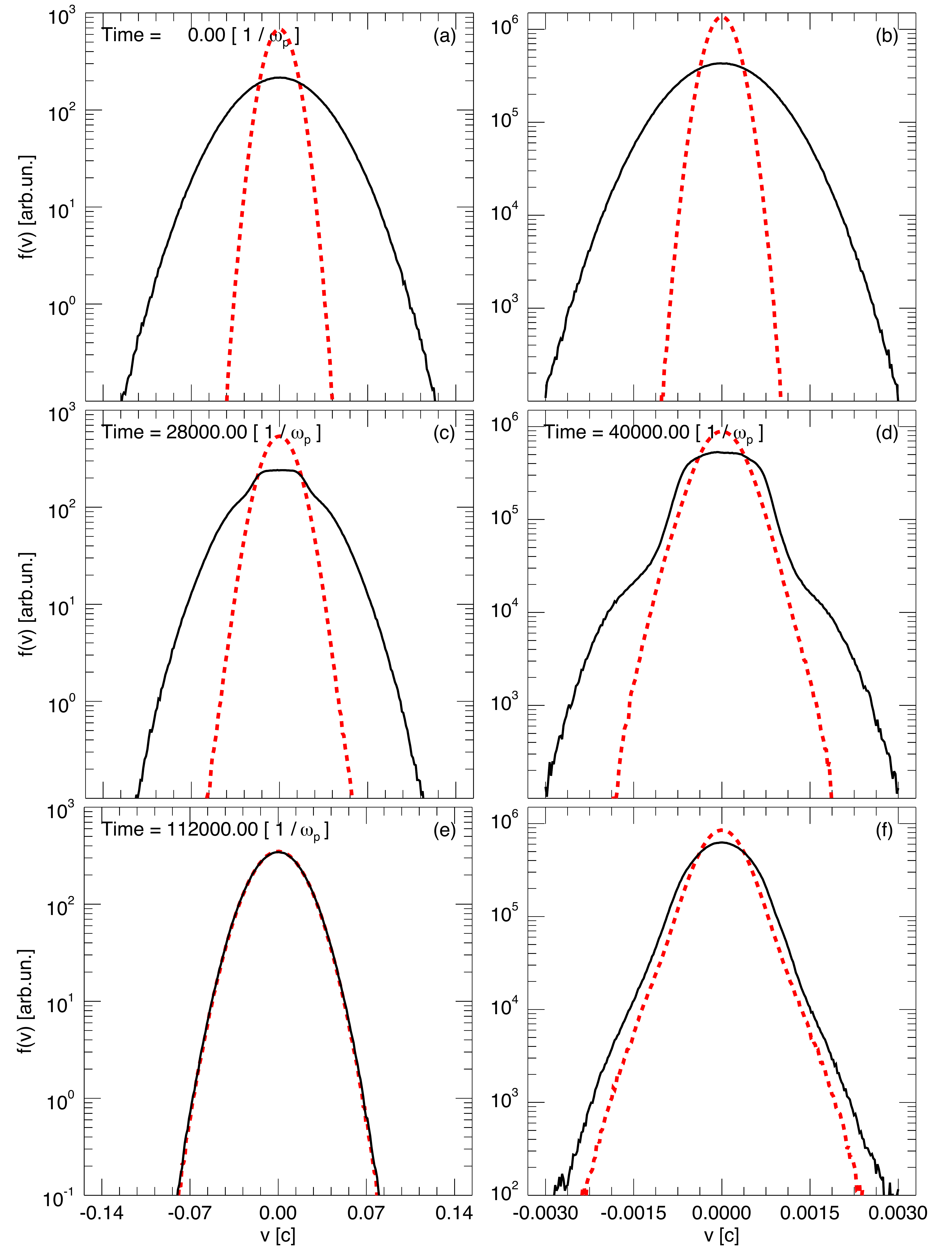}
\caption{Integrated electron (left) and proton (right) VDFs for the PFHI + EFHI case 4 at $t = 0$ (a, b), $28000$ (c), $40000$ (d), $112000$ (e, f). The black solid lines represent the parallel VDF, while the red dashed lines represent the perpendicular VDF.}\label{fig.5}
\end{figure}

To check in detail how particles interact with the generated electromagnetic waves in the PFHI + EFHI  case 4, we present the VDFs for electrons and protons integrated over the parallel and perpendicular directions (Figure \ref{fig.5}). In Figure \ref{fig.5}(a)-(b) the initial anisotropic distributions are shown. At $t=28000\, \omega_p^{-1}$ the perpendicular electron distribution function remains Maxwellian whereas the parallel VDF becomes a flat-top distribution function (Figure \ref{fig.5}(c)). Via the process of scattering on the electromagnetic waves generated due to the instability, the proton distribution function gains suprathermal tails in the direction parallel to the background magnetic field (Figure \ref{fig.5}(d)).  
This represents a signature of a preferential resonant interaction. Finally, while the electrons at the end of the simulation show perfect isotropy with the distribution functions along the parallel and perpendicular directions being essentially identical (Figure \ref{fig.5}(e)), the protons are still slightly anisotropic, with the thermal spread in the parallel direction still larger than the thermal spread in the perpendicular direction (Figure \ref{fig.5}(f)).

\begin{figure}[] 
\centering
\includegraphics[scale=0.42]{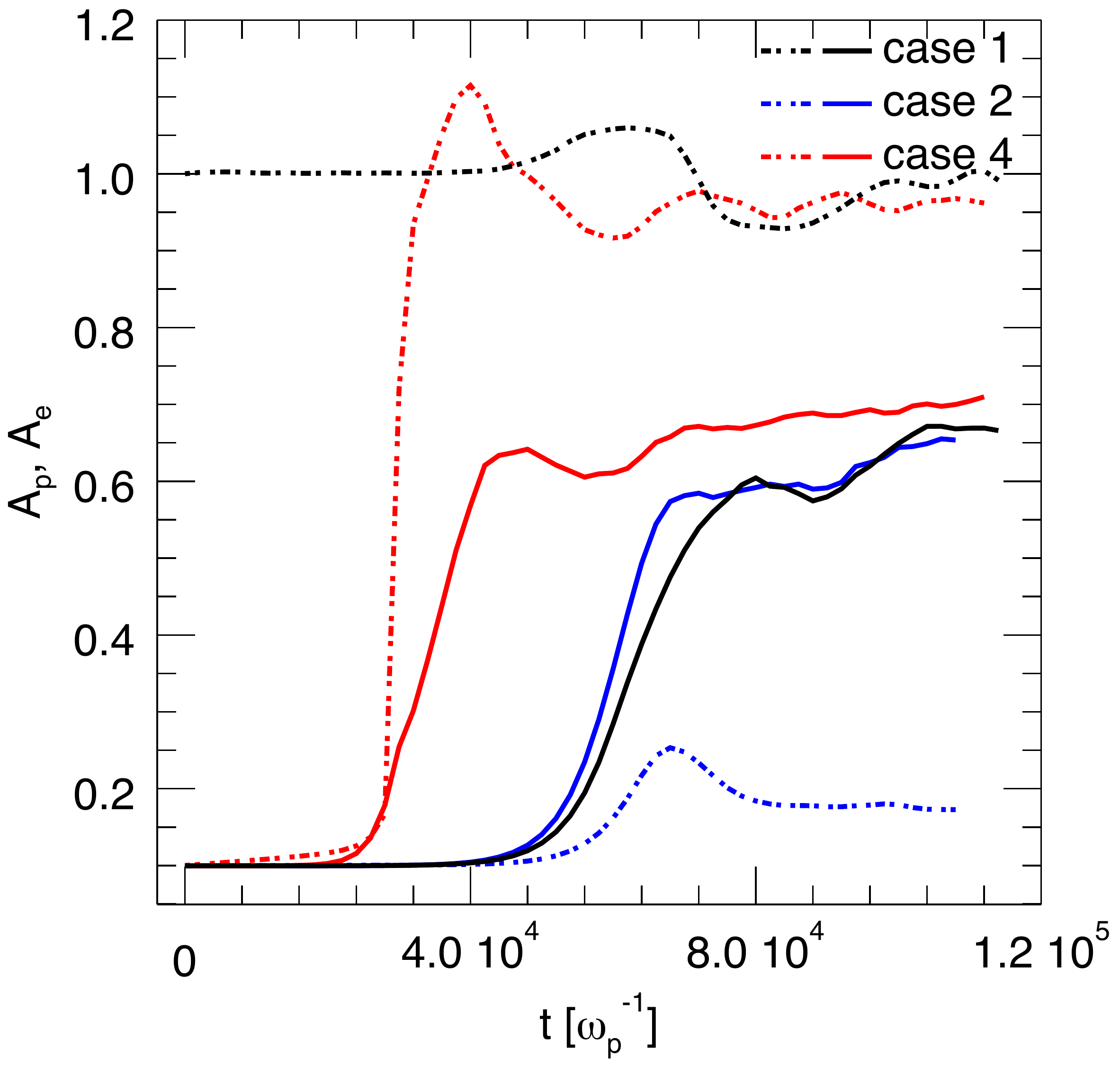}
\caption{Time evolution of proton (solid lines) and electron (dashed-dotted lines) temperature anisotropies for cases 1 (black), 2 (blue) and 4 (red) reported in Table \ref{table:nonlin}.}\label{fig.10}
\end{figure}

Figure \ref{fig.10} shows the evolution of proton and electron temperature anisotropies, $A_p$ and $A_e$, respectively for cases 1, 2 and 4 reported in Table \ref{table:nonlin}. Both species, when they are unstable with respect to the firehose instability, are subjected to parallel cooling and perpendicular heating. The result is a reduction of the initial anisotropy with time and hence the corresponding increase of $A_p$ and $A_e$.
Evolution of the proton temperature anisotropy for the PFHI is essentially the same for cases 1 and 2. This is due to the fact that the electrons, in both the cases, are stable with respect to the parallel electron firehose. The parallel proton firehose instability affects only marginally the electrons when they are stable with respect to the electron firehose instability. Indeed, in these two cases the electron temperature anisotropy $A_e$ presents only small oscillations around the stable initial condition. This means that the proton firehose instability can be properly described also via hybrid simulations if electrons are not subject to the electron firehose instability.

For the PFHI + EFHI (case 4), characterized by the temperature anisotropy of both species and high electron plasma beta, one can notice that the electrons reach the isotropic state on a shorter time scale than protons. In this case, the interplay between electron and proton anisotropies modifies the onset of the proton firehose instability, which leads to an earlier isotropization of the proton distribution functions, if compared with cases 1 and 2.
Moreover, PFHI + EFHI case 4 exhibits the final proton temperature anisotropy that is slightly lower compared to that obtained for the two PFHI cases 1 and 2, where only the protons are unstable with respect to the firehose instability. This result is in agreement with the works by \citet{Yoon2017b} and \citet{Yoon2019}, where the coupling between electrons and protons is considered necessary to explain the broad distribution of particles near isotropic conditions in the solar wind. In addition, we note that the final value of $A_p$ is around $0.7$, which is in agreement with the most probable level of proton anisotropy observed in the solar wind at 1 AU \citep{kasper2003}.

\begin{figure}[] 
\centering
\includegraphics[scale=0.3]{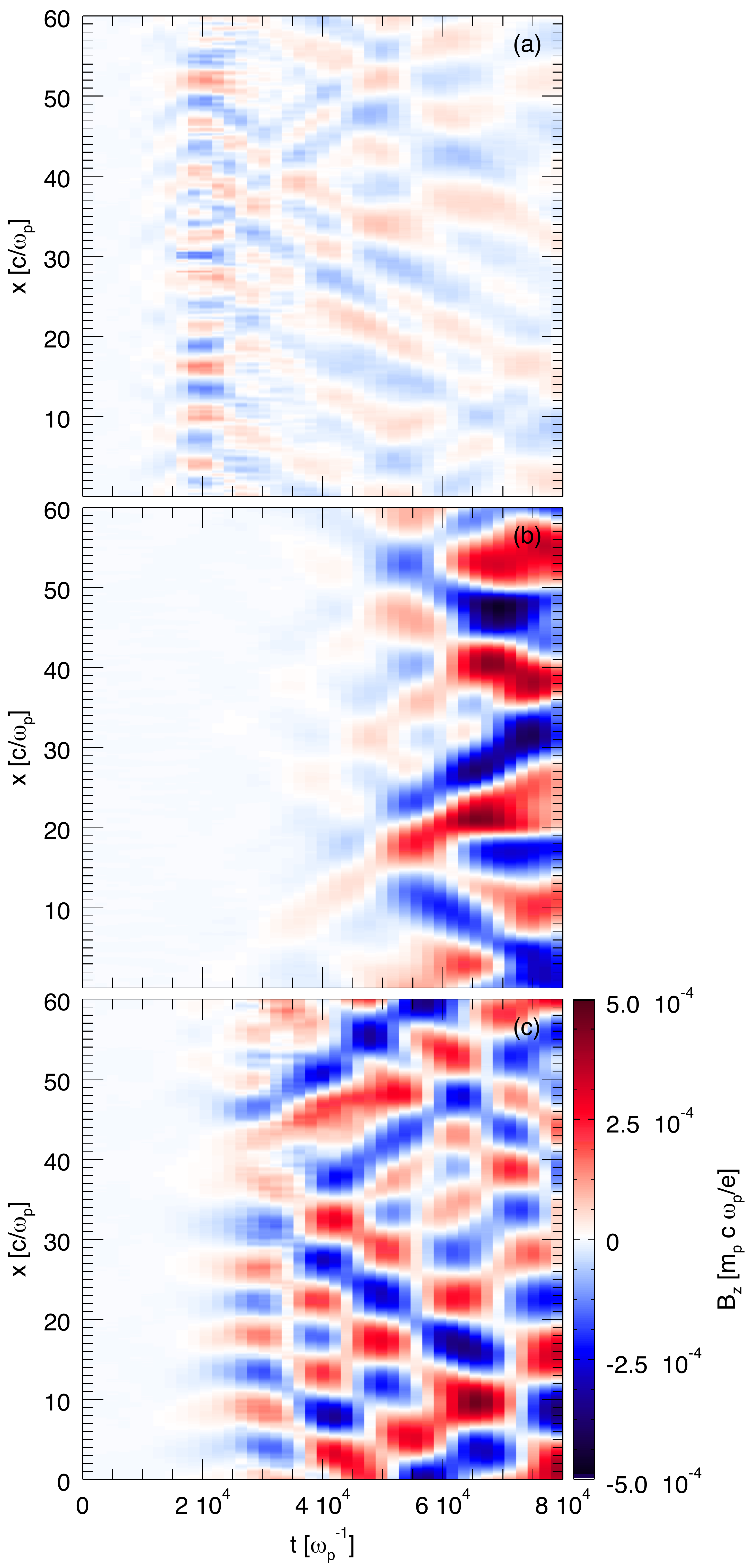}
\caption{Evolution of the out-of-the-plane component of the magnetic field $B_z$ for the EFHI case 3 (a), PFHI case 1 (b), and PFHI + EFHI case 4 (c).}\label{fig.2}
\end{figure}

Figure \ref{fig.2} shows the evolution of the out-of-the-plane magnetic field $B_z$ for the three analyzed cases. It demonstrates 
the development of the wave fluctuations and the increase of their amplitude with time as particle isotropize due to the firehose instability.
The more rapid development of the instability in the PFHI + EFHI  case with respect to the PFHI case can be easily noticed. 
Comparing the three panels of Figure \ref{fig.2}, it is possible to identify characteristic time scales for the rise of the magnetic fluctuations. The instability time scale depends on the presence or absence of the interplay between proton and electron anisotropies.
In case of anisotropic electrons and isotropic protons (EFHI case) the instability develops relatively fast, and the magnetic fluctuations reach the peak amplitude around $20000$ $\omega_p^{-1}$. When bi-Maxwellian distributions are initialized for both protons and electrons, the resulting kinetic instability takes more time to develop with respect to the EFHI case, but it appears almost twice faster than in the PFHI case. In all cases it is clear from Figure \ref{fig.2} that the parallel firehose instability is characterized by propagating modes ($\omega_r \ne 0$).

\section{Comparison of simulation results with the linear theory}\label{sec.3}

In this section we perform a comparison between the linear theory and our PIC simulations.

The linear stability analysis is based on the linearized Vlasov-Maxwell dispersion relation for the wave propagation parallel to the background magnetic field in a homogeneous and bi-Maxwellian electron-proton plasma.  
Linear theory can predict the fastest growing plasma modes and describe their dispersive characteristics such as growth rates, wave frequency and instability thresholds. However, linear theory cannot describe neither the interaction between the different plasma modes nor the energy transfer between plasma species.
Only PIC and hybrid simulations contain the full description of the nonlinear physics.

For the firehose unstable modes the normalized dispersion relation is as follows \citep{Shaaban2017}:
\begin{equation}
     \begin{aligned}
\tilde{k}^{2}=&\mu \left({A_e}-1\right)+\mu \frac{{A_e}\left( \tilde{\omega}- \mu\right)+ \mu}{\tilde{k}\sqrt{\mu \beta_{e,\parallel}}}Z_{e}\left( \frac{\tilde{\omega}- \mu }{\tilde{k}\sqrt{\mu \beta _{e,\parallel}}}\right)\\
&+A_{p}-1+\frac{ A_{p}\left(\tilde{\omega}+ 1\right)- 1}{\tilde{k}\sqrt{\beta _{p,\parallel}}}Z_{p}\left( \frac{\tilde{\omega}+ 1}{\tilde{k}\sqrt{\beta_{p,\parallel}}}\right),
    \label{dis}   
     \end{aligned}     
\end{equation}
where $\tilde{k}=~kc/\omega_{p}$ is the normalized wave-number $k$, $\tilde{\omega}=\omega /\Omega _{p}$ and  
\begin{equation}
Z_{j}\left( \xi _{j}^{\pm }\right) =\frac{1}{\pi ^{1/2}}\int_{-\infty
}^{\infty }\frac{\exp \left( -x^{2}\right) }{x-\xi _{j}^{\pm }}dt \label{e5}
\end{equation}
is the plasma dispersion function \citep{Fried1961} of argument $\xi _{j}^{\pm }= \left(\omega \pm \Omega _{j}\right)/\left(ku_{j,\parallel}\right)$, with the imaginary part $\Im \left( \xi _{j}^{\pm }\right) >0$.

We solve the dispersion relation (\ref{dis}) numerically for the same four sets of plasma parameters used in our simulations and listed in Table \ref{table:nonlin} (see Section \ref{seccc.2}).

\begin{figure}[]
\centering
 \includegraphics[scale=0.55, trim={0 1cm 0 0.8cm}, clip]{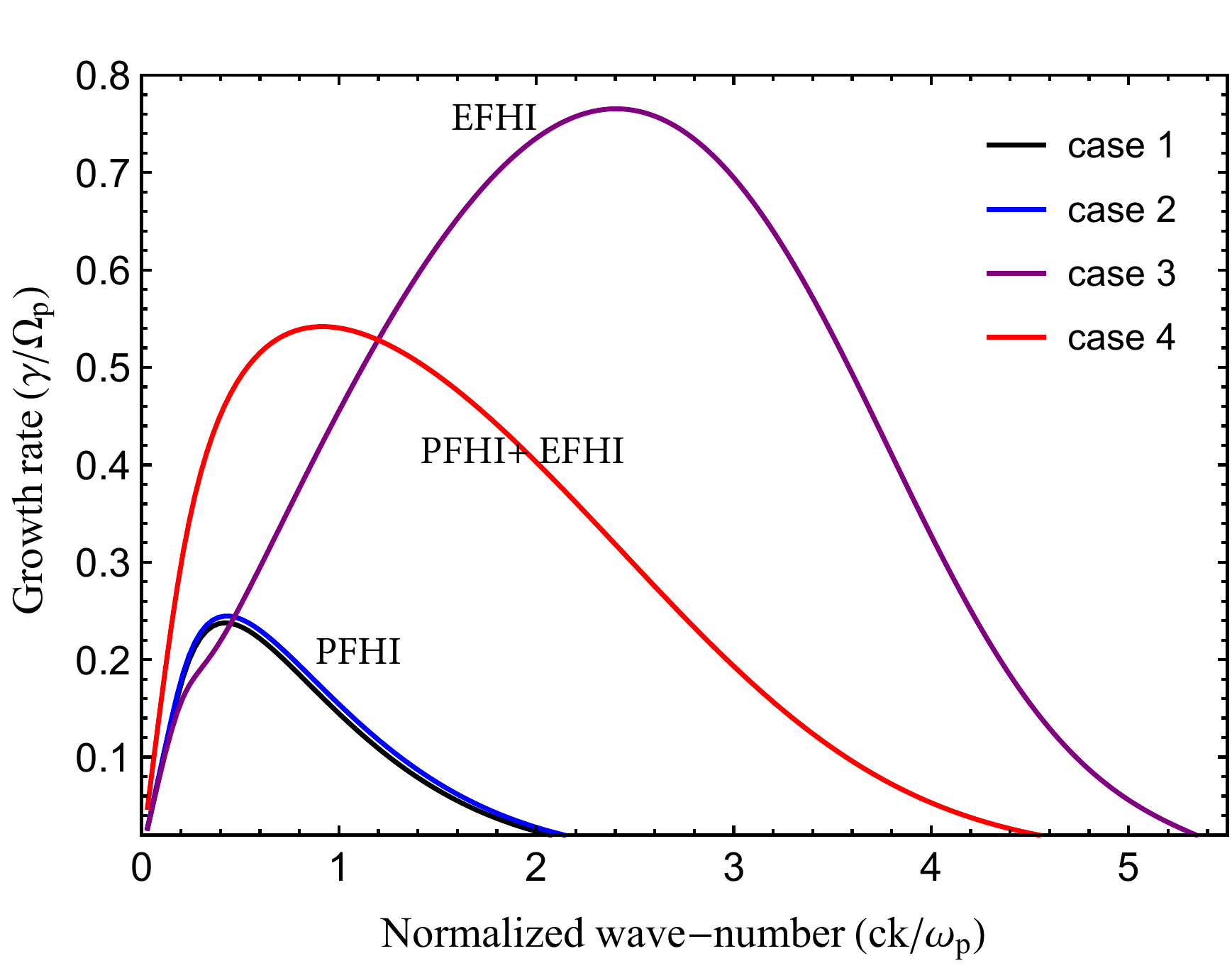}
 \includegraphics[scale=0.55, trim={0 0 0 0.8cm}, clip]{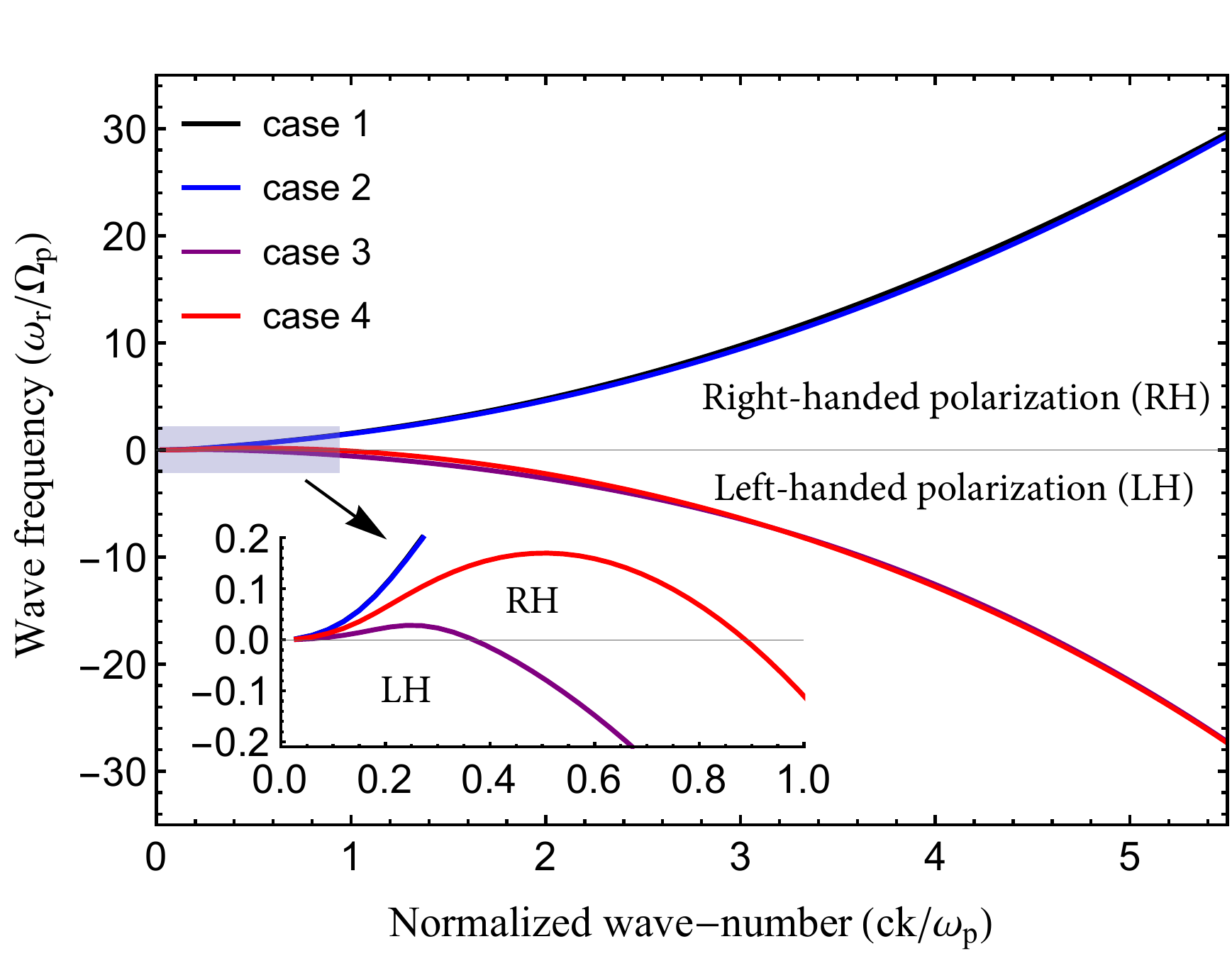}
 \caption{Growth rates of the firehose instability (top panel) and wave frequencies of the analyzed unstable modes (bottom panel) as a function of the normalized wavenumber for cases 1 -- 4. The inset in the bottom panel magnifies the gray-shaded area.} \label{f2.pdf}
\end{figure}

Figure \ref{f2.pdf} displays the unstable solutions of the firehose modes shown by the growth rates in the top panel and the wave frequencies in the bottom panel as functions of the normalized wavenumber. For case 1 the contribution of electrons is minimized by considering them isotropic, and therefore only the proton firehose instability is excited by the anisotropic protons with $A_p=0.1$, see black line. The growth rate and wave frequency of the unstable proton firehose mode obtained for case 1 serves as a reference for the rest of the results that are obtained for the other cases. The presence of anisotropic electrons can markedly alter the dispersive characteristics of the unstable proton firehose modes \citep{Michno2014, Maneva2016, Shaaban2017}. Therefore, in cases 2 and 4 (unstable proton firehose modes under the influence of the electron anisotropy for two different electron plasma betas), it is clear that anisotropic electrons with low plasma beta $\beta_{e,\parallel}=0.1$ have a negligible effect on the PFHI. For cases 1 and 2, only the PFHI is excited and the wave frequency is right-hand (RH) polarized (which we define as the frequency with positive $\tilde{\omega}_r$). For sufficiently large electron plasma beta, i.e. $\beta_{e,\parallel}=4.0$, the left-hand (LH) EFHI is excited at high normalized wavenumbers $\tilde{k}>1$ by anisotropic electrons with $A_e=0.1$. This leads also to the enhancement of the PFHI at low normalized wavenumbers $\tilde{k}<1$. The wave frequency plot in the bottom panel confirms the presence of both instabilities. The growing wave is RH-polarized at small wavenumbers and LH-polarized at high wavenumbers (the red curve changes the sign at at $\tilde{k} \sim 1$). In case 3 we obtain the unstable solutions of the EFH modes. With isotropic protons, the EFHI exhibits a larger growth rate and extends to larger wavenumbers compared to those obtained for case 4. Straightforward comparison between the results obtained for cases 3 and 4 shows clearly that anisotropic protons with $A_p<1.0$ have inhibiting effects on the EFHI, decreasing both growth rates and unstable wavenumbers. The corresponding wave frequency is negative along the bulk of the solution including the part of the normalized wavenumber space around the maximum of the growth rate. This confirms the LH polarization of the unstable EFH mode.

\begin{figure}[] 
\centering
\includegraphics[scale=0.42]{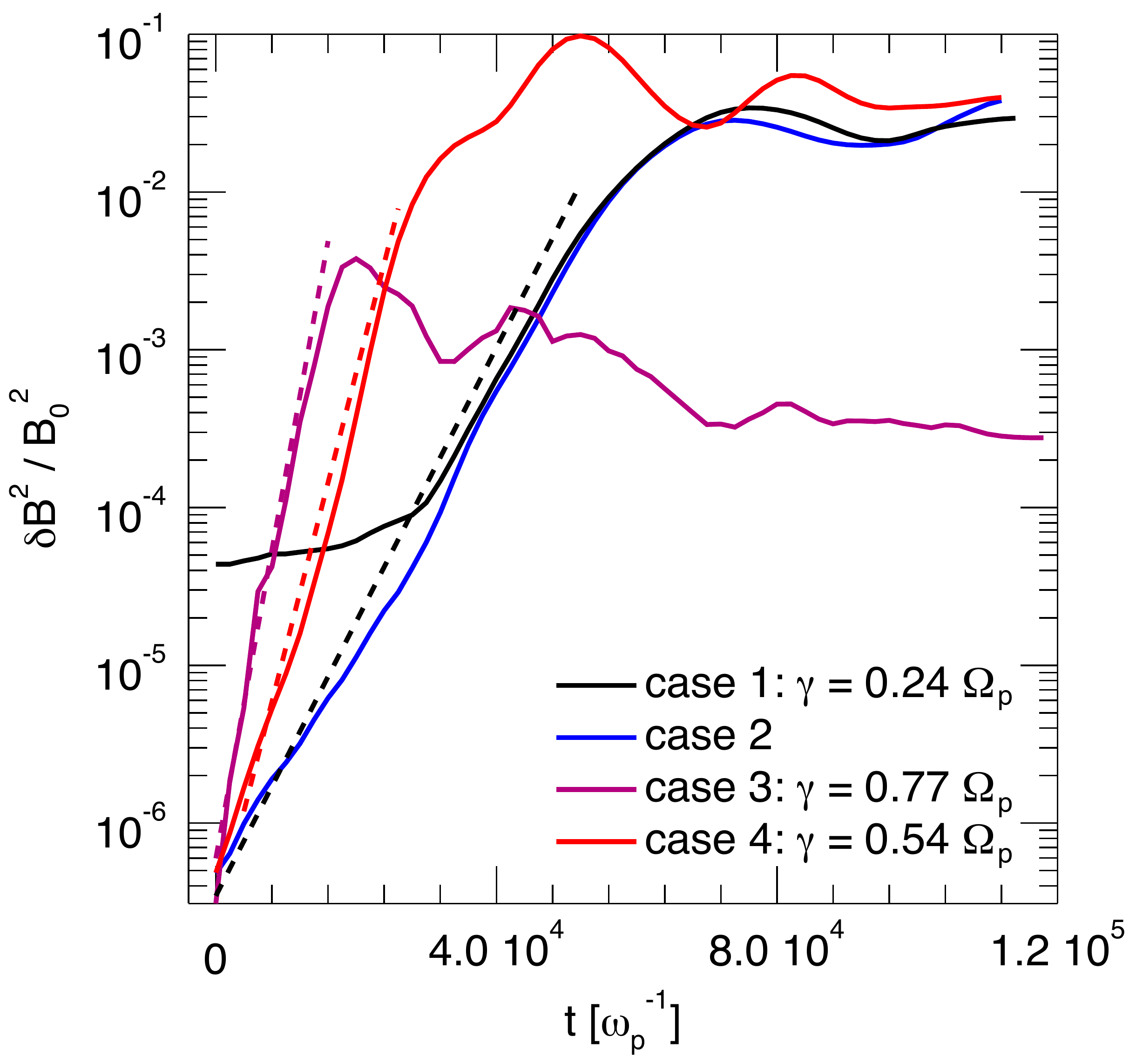}
\caption{Time evolution of the magnetic energy fluctuations for the four runs of our numerical simulations corresponding to the four cases reported in Table \ref{table:nonlin} (solid lines) and comparisons with the corresponding theoretical growth (dashed lines).}\label{fig.1}
\end{figure}

In order to compare our PIC simulations and the linear theory, we show in Figure \ref{fig.1} the fluctuations of the magnetic energy, normalized to the initial magnetic field $\bm{B}_0$, and compare their growth with relative growth rates obtained from the linear theory, for the four cases listed in Table \ref{table:nonlin}.
In all the simulations the system evolves from an initially unstable regime to the state of marginal stability, due to the interaction between waves and particles. We note that the very low level of magnetic fluctuations close to $t=0$, before the development of the instabilities, depends on whether or not electrons are initially isotropic. This is linked to the fact that we do not use any driving except the initial temperature anisotropies (i.e. we do not seed the instability). Indeed, in all our simulations the instability grows starting from the numerical noise. Different level of initial magnetic fluctuations does not affect the linear growth rate of the instability, as expected.
The magnetic fluctuations grow with time as a signature of an exchange of energy between the magnetic field 
and the particles. The saturation is reached once the anisotropy in the electron or/and proton velocity distributions has almost entirely disappeared.

Time evolution of the magnetic energy fluctuations for the PFHI, is almost identical in cases 1 and 2, as indicated in Figure \ref{fig.1} with the black and the blue solid lines respectively. 
The magnetic energy fluctuations in both linear and nonlinear regimes of the instability are similar.
The theoretical growth rate ($\gamma$) for these two sets of unstable conditions is shown in Figure \ref{fig.1} as a dashed black line corresponding to $\gamma = 0.24\, \Omega_{p}$. Both cases lead to the same evolution of the magnetic energy fluctuations, as for small values of the electron plasma  $\beta$ the electron firehose does not develop and the electron temperature anisotropy has no effect on the growth rate of the parallel proton firehose instability. 

For the EFHI (case 3), the evolution of the magnetic energy fluctuations is shown in Figure \ref{fig.1} with the purple solid line. The maximum growth rate in this case is around three times higher than the PFHI growth rate. Indeed, the maximum growth rate according to the linear theory is $\gamma = 0.77\, \Omega_{p}$  and is indicated in Figure \ref{fig.1} with the dashed purple line.

The main finding of this work is related to the evolution of the magnetic energy fluctuations for the PFHI + EFHI  (case 4), displayed in Figure \ref{fig.1} with a red solid line. In this case, as shown above, the interplay between the electron and proton anisotropies significantly modifies the development of the instability. 
The growth rate of the parallel proton firehose instability is enhanced by the electron anisotropy in comparison with the pure proton instability (dashed black line in Figure \ref{fig.1}). Indeed, this case is characterized by the linear growth rate $\gamma = 0.54\, \Omega_{p}$ (dashed red line). 

\begin{figure}[] 
\centering
\includegraphics[scale=0.3]{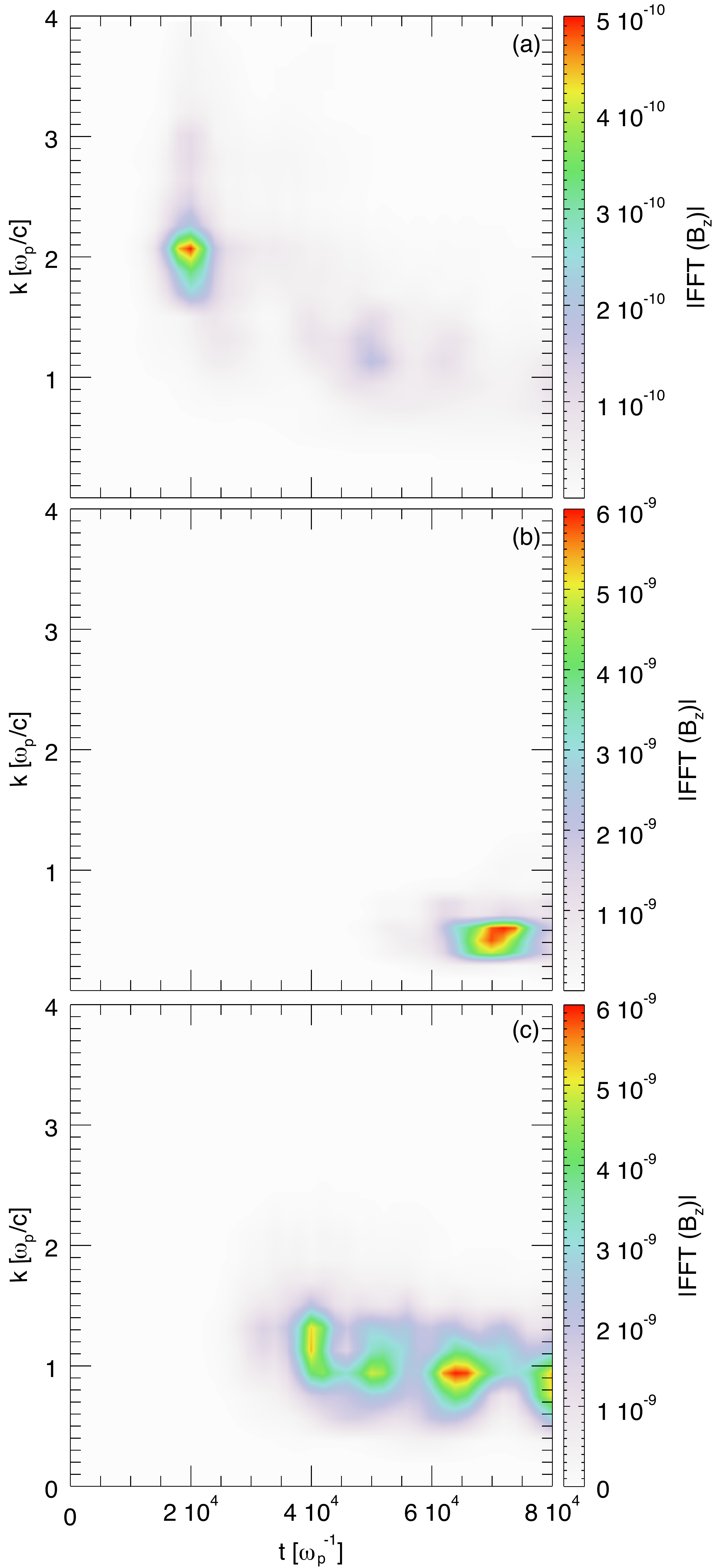}
\caption{Evolution of the normalized wavenumber $k$ for the EFHI case 3 (a), PFHI case 1 (b), and PFHI + EFHI case 4 (c).}\label{fig.11}
\end{figure}

In order to verify that our simulations properly captured the linear stage of electron and proton parallel firehose instabilities and to have an additional comparison with the linear theory, Figure \ref{fig.11} shows the temporal evolution of the normalized wavenumber $k$ for cases 1, 3 and 4, obtained as fast Fourier transforms in space of the out-of-the-plane component of the magnetic field $B_z$. In Figure \ref{fig.11} one can remark the more rapid development of the electron firehose instability and its extension to higher wavenumbers compared to the other two cases. Furthermore, comparing the PFHI + EFHI case 4 and the PFHI case 1 it is possible to notice the earlier development of the firehose instability in the presence of the electron temperature anisotropy and the higher wavenumbers which characterize case 4 in comparison with case 1.
These results are in qualitative agreement with those obtained via linear theory as shown in the top panel of Figure \ref{f2.pdf}. 

Figure \ref{fig.11} (c) shows also that other modes develop after the saturation of the fastest growing mode around $t=40000\ \omega_p^{-1}$ (cf. Figure \ref{fig.1}). In particular, we note that a very energetic mode with $ck/\omega_p \sim 1$ grows around $t=65000\ \omega_p^{-1}$. At this time, the evolution of the oscillations of the magnetic field energy displays an increment (see red solid curve in Figure \ref{fig.1}).

\section{A test of oblique cases}\label{sec.4}
Three-dimensional (3D) and two-dimensional (2D) full PIC simulations are required in order to obtain a more complete picture of the processes that take place in the coupled dynamics of parallel and oblique firehose instabilities, when the kinetic description for both species is retained. This kind of problems can hardly be modeled in multi-dimensional simulations with real mass ratio due to extremely challenging computational effort. In an attempt to test the coupling between electrons and protons in oblique cases, we performed 1D simulations at oblique angles with respect to the background magnetic field, which allow the development of oblique electron and proton firehose instabilities. Using the same parameters as described in Sections \ref{sec.2} and \ref{seccc.2}, we present a comparison between the PFHI case 1 and the PFHI + EFHI case 4 at $\theta = 0^{\circ}$, $\theta = 35^{\circ}$, $\theta = 45^{\circ}$ and $\theta = 60^{\circ}$, where $\theta$ is the angle between the directions of the magnetic field and the $x$-axis of the simulation box.

\begin{figure}[] 
\centering
\includegraphics[scale=0.42]{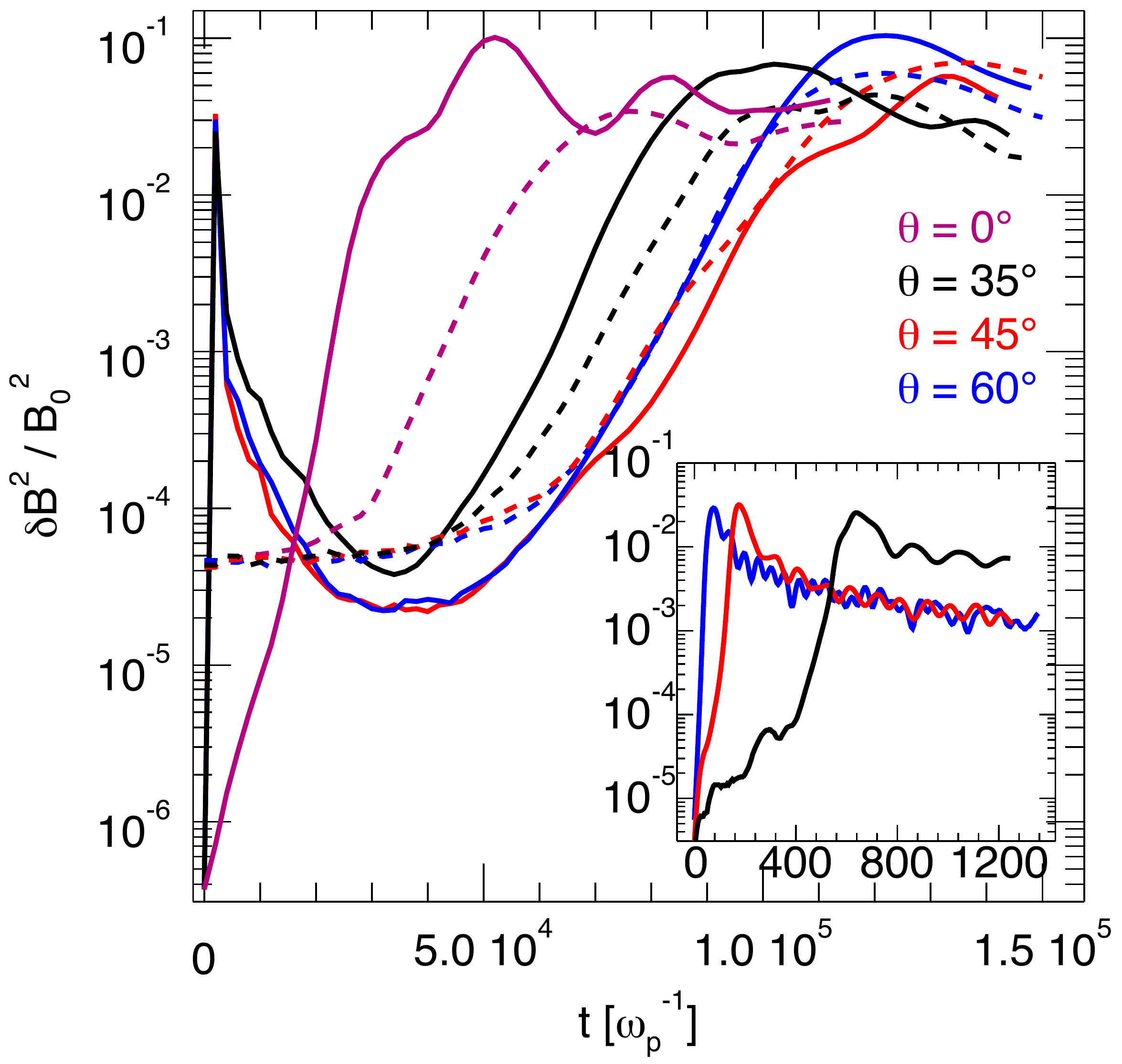}
\caption{Evolution of the magnetic energy for the PFHI + EFHI case 4 (solid lines) and the PFHI case 1 (dashed lines) for the simulations at propagation angles of $0^{\circ}$ (purple), $35^{\circ}$ (black), $45^{\circ}$ (red) and $60^{\circ}$ (blue). The inset shows the detailed evolution between 0 and 1200 $\omega_p^{-1}$ for the PFHI + EFHI case 4 at oblique angles with respect to $B_0$.}\label{fig.12}
\end{figure}

Figure \ref{fig.12} shows the temporal evolution of the magnetic energies for eight different runs, corresponding to PFHI + EFHI case 4 and PFHI case 1 for different orientations of the simulation box with respect to the initial background magnetic field. In this way we can compare the growth rates of parallel and oblique proton firehose instabilities and reveal how the presence of anisotropic electrons with $A_e < 1$ affects the development of the proton firehose instability at each propagation angle. Each simulation starts from an unstable condition and the instability grows until the magnetic energy saturation, which corresponds to the quasi-isotropization of the proton VDF. In case 4, at oblique angles, the initial peak corresponds to the electron firehose and the later peak corresponds to the proton firehose.

\begin{figure}[] 
\centering
\includegraphics[scale=0.42]{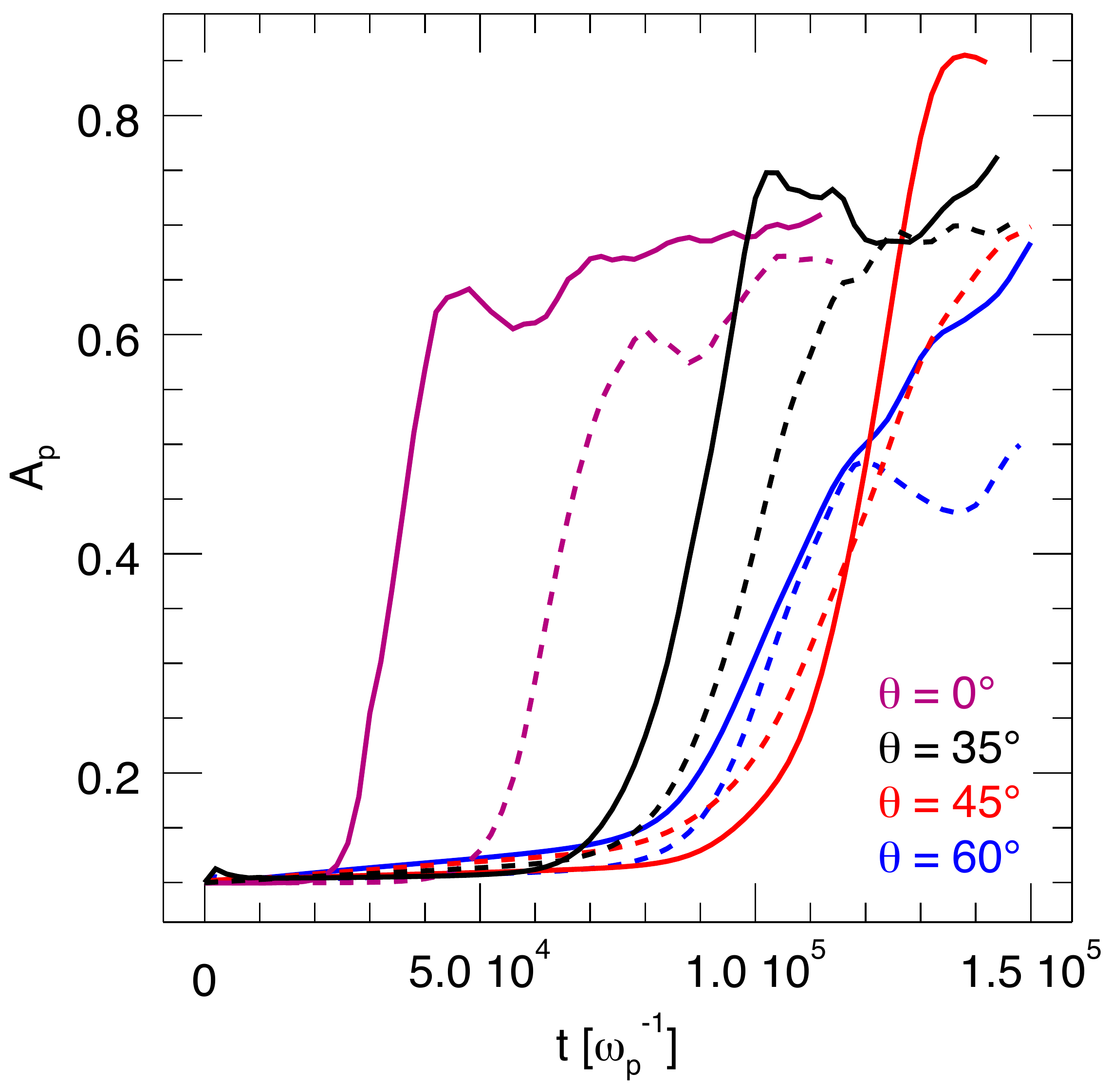}
\caption{Evolution of the proton temperature anisotropy in the PFHI + EFHI case 4 (solid lines) and PFHI case 1 (dashed lines) for the simulations at propagation angles of $0^{\circ}$ (purple), $35^{\circ}$ (black), $45^{\circ}$ (red) and $60^{\circ}$ (blue).}\label{fig.13}
\end{figure}

Our result reported in Section \ref{seccc.2}, calculated at $\theta = 0^{\circ} $, is confirmed to be valid also at oblique angles of propagation. Indeed, also for $\theta= 35^{\circ}$ (black lines) and $\theta= 45^{\circ}$ (red lines) the case 4, characterized by the anisotropy of both electrons and protons with $A_j < 1$ (solid lines), presents a higher growth rate and a faster magnetic energy saturation than the case where only the protons are anisotropic (dashed lines). The difference in the growth rates becomes less evident at higher angles of propagation (blue curves in Figure \ref{fig.12}). However, the impact of the electrons on the development of proton firehose instabilities is clearly present in the temporal evolution of proton temperature anisotropies. Indeed, the final proton temperature anisotropy is lower ($A_p$ is higher) in PFHI + EFHI case 4 compared to the PFHI case 1 at all the different propagation angles analyzed (see Figure \ref{fig.13}).

We confirm that the parallel proton firehose instability has a lower threshold and a higher growth rate as compared to the oblique proton firehose instability \citep{HellingerMatsumoto2000, Kasper2002, Hunana2017, Hunana2019}.Although for these sets of parameters the dominant oblique electron firehose instability saturates on fast, electron scales and the electron temperature anisotropy is reduced much earlier than the proton temperature anisotropy, the interplay between the two species is still present and clearly affects the development of the proton oblique firehose instability.

Figure \ref{fig.13} shows the temporal evolution of proton temperature anisotropies in the PFHI + EFHI case 4 and PFHI case 1 at the same four angles $\theta$ listed above. In all the cases analyzed, the proton VDFs react to the increase of the magnetic energy, reducing their initial anisotropy until a stable quasi-isotropic state is reached. For each angle of propagation, the isotropization is faster when the electrons are also unstable with respect to the oblique firehose instability (solid lines), as compared to the case 1 where the electrons are initially isotropic (dashed lines). Moreover, the proton isotropization is faster in the parallel firehose instability than in the oblique firehose instability, due to the higher growth rate of the former. The electron temperature anisotropies are not shown in Figure \ref{fig.13} for the sake of clarity, since the electron isotropization in the PFHI + EFHI case 4 occurs on fast scales, while they remain isotropic in the PFHI case 1. However, it has to be pointed out that the interplay of instabilities at different angles is not captured in our 1D setup and has to be studied in more realistic 2D and 3D simulations.

\section{Discussion and conclusions}\label{sec.5}
In order to explain a low level of anisotropy recorded for both electrons and protons around 1 AU and beyond \citep{Matteini2007}, it has been suggested that electromagnetic micro-instabilities could play a fundamental role in limiting the difference among the diagonal elements of the pressure tensor \citep[see e.g. the review by][]{Marsch2006}. 
Indeed, plasma electromagnetic instabilities in the solar wind are processes that regulate and limit the anisotropies in the particle velocity space, so the observed velocity distribution functions for both protons and electrons are at the margin of stability \citep{Stverak2008}, \citep{Kasper2016}. The firehose instability is thought to have a significant role in regulating electron and proton temperature anisotropies in the solar wind.
Verifying this hypothesis requires fully kinetic simulations of plasma conditions in the solar wind. 

We have studied the parallel firehose instability for protons and electrons via 1D fully kinetic simulations performed with ECsim, a semi-implicit PIC code.
The main focus of this work is the interplay between anisotropic protons and electrons with $A_j < 1$. Our  non-linear fully kinetic simulations demonstrate that the electron temperature anisotropy,  $A_e < 1$, leads to a faster development of the proton parallel firehose instability, as suggested by the linear theory \citep{Michno2014, Shaaban2017}.

The presence of an electron temperature anisotropy has a limited impact on the level of the magnetic field fluctuations at saturation. The case characterized by the temperature anisotropy of both species and high electron plasma beta (PFHI + EFHI case 4) exhibits the final proton temperature anisotropy which is slightly lower compared to that obtained for the pure proton firehose instability (PFHI case 1), where only the protons are anisotropic. The final value of $A_p$ around $0.7$ is in agreement with the most probable level of proton anisotropy observed in the solar wind at 1 AU \citep{kasper2003}. The proton firehose instability affects only marginally the electrons when they are stable with respect to the firehose instability. Indeed, in these cases the electron temperature anisotropy presents only small oscillations around the stable initial conditions.

The use of a semi-implicit PIC code has given us the opportunity to use a large simulation box that does not limit the number of unstable modes (more than 20 wave lengths of the fastest growing mode fit into the box). In addition we use a realistic mass ratio and a frequency ratio $\omega_{e} / \Omega_{e} = 63.24$, that is relevant for the solar wind \citep{Tong2019}. 
The results of our fully kinetic simulations are in good agreement with the ones obtained with the linear theory.
This is a confirmation that ECsim models the physics correctly even at relatively coarse resolutions.

An important limitation of our work is that it does not take into account the competition between parallel and oblique proton firehose instability. Our simulation captures only the parallel character of the analyzed unstable mode. We note that it is still an open question which instability limits the observed proton temperature anisotropies. 
On the one hand, hybrid simulations reveal that both parallel and oblique proton firehose instabilities are relevant in the solar wind context and show comparable growth rates for a wide range of parameters \citep{HellingerMatsumoto2000}. 
On the other hand, Wind/SWE observations \citep{Kasper2002} indicate that there exists a phenomenological constraint on the proton temperature anisotropy which is compatible with the threshold of the parallel firehose instability. 
Despite taking into account only the parallel instability, our analysis is relevant for understanding the consequences of the combination of ion and electron temperature anisotropies in the linear and non-linear regimes of the instability.
The study of the oblique instabilities goes beyond the scope of this work and will be a subject of further investigations.
More realistic and more computationally demanding case to describe the full spectrum of firehose instabilities goes beyond the scope of this work as two- or maybe even three-dimensional simulations \citep[see][]{Howes2015} would be necessary.

This work focuses on the periodic proton firehose instability and aims at understanding how it is affected by an electron temperature anisotropy. Even if the oblique electron firehose instability has generally a lower threshold and higher growth rate than the parallel electron firehose instability, numerical simulations by \citet{Camporeale2008} demonstrate that quasi-parallel, propagating modes of the electron firehose instability are excited in the non-linear stage of the instability. In particular, \citet{Camporeale2008} show the shift of the most unstable modes toward smaller angles of propagation until the time when the oblique, non-propagating modes with the highest growth rates saturate and the electron firehose instability becomes driven by modes at any angle of propagation. \citet{Camporeale2008} state that over a certain period of time even exactly parallel modes (propagation angle of 0$^{\circ}$) dominate. We note that, as mentioned in Section 1, the three properties - being non-resonant (for electrons), parallel and propagating - are often used as definitions of the parallel firehose instability \citep{LiandHabbal2000,GaryandNashimura2003,Camporeale2008}, and all are well captured in our simulations. 

Another limitation of our simulations is that we assumed initial bi-Maxwellian distributions for protons and electrons.
The presence of suprathermal populations, which are ubiquitous in space plasmas \citep{Marsch2006, Stverak2008, Bercic2019}, also affects the instability thresholds and may alter the competition between parallel and oblique firehose modes \citep{Lazar2011, Astfalk2016, Lopez2019, Shaaban2019c}.
Moreover, linear and non-linear investigations of the properties of parallel and oblique firehose instabilities have determined that the growth rates of these two instabilities are strongly affected by the presence of alpha particles and by their properties \citep{Dasso2003, Hellinger2006,Ofman2019}. 
More complex structures of particle velocity distribution functions, as well as more realistic plasma composition, introduce other sources of free energy for kinetic instabilities, which will be investigated in future work.

\acknowledgments
This work was supported by a PhD grant awarded by the Royal Observatory of Belgium to one of the authors (A. M.).
These simulations were performed on the supercomputers SuperMUC (LRZ) and Marconi (CINECA) under PRACE and HPC-Europa3 allocations.
Authors thank M. E. Innocenti for helpful discussions.
A. N. Z. thanks the European Space Agengy (ESA) and the Belgian Federal Science Policy Office (BELSPO) for their support in the framework of the PRODEX Programme. 
S. M. S. acknowledges support by a FWO Postdoctoral Fellowship (Grant No. 12Z6218N).

\bibliography{main}

\listofchanges
\end{document}